\documentstyle[preprint,epsf,aps]{revtex}

\preprint{IMSc/98/06/26}
\begin{document}
\draft 
\title{One pion events by atmospheric neutrinos:
A three flavor analysis}
\author
{K. R. S. Balaji, G. Rajasekaran }
\address
{Institute of Mathematical Sciences, Chennai 600 113, India.}
\author
{S. Uma Sankar}  
\address
{Department of Physics, I.I.T. , Powai, Mumbai 400076, India}
\date{\today}
\maketitle

\begin{abstract}
We study the one-pion events produced via neutral current (NC) and 
charged current (CC) interactions by the atmospheric neutrinos. We 
analyze the ratios of these events in the framework of oscillations 
between three neutrino flavors. The ratios of the CC events induced by $\nu_e$ 
to that of the NC events and a similar ratio defined with $\nu_\mu$ help us in 
distinguishing the different regions of the neutrino parameter space.
\end{abstract}
\vspace{0.5cm}
\narrowtext

\section{INTRODUCTION}

It has been more than a decade since the atmospheric neutrino anomaly
has been observed by the water Cerenkov detectors IMB \cite{imb} and
Kamiokande \cite{hirata,fukuda}. Eventhough conventional detectors
Frejus \cite{frejus} and NUSEX \cite{nusex} have not observed this 
effect, a recent experiment SOUDAN-II did indeed see a deficit
\cite{soudan}. Recent results from Superkamiokande confirmed the 
earlier results including the zenith angle dependence of the deficit
in the multi-GeV data \cite{superk}. The results of all these   
experiments are presented in the form of the double ratio  
\begin{equation}
R = \frac{\left(\frac{N_{\nu_{\mu}}}{N_{\nu_e}}\right)_{obs}}
{\left(\frac{N_{\nu_{\mu}}}{N_{\nu_e}}\right)_{MC}}
= \frac{r_{obs}}{r_{MC}}. \label{eq:defR}
\end{equation}
The measured value of $R$ for Kamiokande is 
${0.60}^{+0.07}_{-0.06}~\pm ~0.05$ for sub-GeV data (E$\ <$ 
1.33 GeV) and ${0.57}^{+0.08}_{-0.07}~\pm ~0.07$ 
for the multi-GeV data (E$\ >$ 1.33 GeV)
whereas that of 
Superkamiokande is $0.61~\pm ~0.06~\pm ~0.05$ 
for sub-GeV and $0.67~\pm ~0.06~\pm ~0.08$ for the 
multi-GeV data. The 
reason for presenting the result in the form of the double ratio is that 
the theoretical calculations
of $\nu_\mu$ and $\nu_e$ fluxes are subject to large uncertainties
of the order of $30 \%$ \cite{honda,agarwal}.
Hence a comparison of measured neutrino flux with the calculated 
one (either for $\nu_\mu$ or for $\nu_e$) is not particularly useful.
However, the uncertainties in the ratios of the calculated fluxes
are much smaller (less than $10 \%$). Hence a comparison of the 
measured ratio of $\nu_\mu$ flux to $\nu_e$ flux to the calculated 
ratio will yield meaningful information on neutrino properties. 

Neutrino oscillations provide a natural explanation for both the 
overall deficit and the zenith angle dependence \cite{pakvasa,gvdkvl}. 
Kamiokande analyzed
their data in terms of two flavor oscillations between $\nu_\mu
\leftrightarrow \nu_e$ and $\nu_\mu \leftrightarrow \nu_\tau$. In
both cases they found allowed regions of parameter space, with the 
mass-squared difference around $\Delta m^2 \simeq 0.02 \ eV^2$ and 
the mixing angle near its maximal value of $\pi/4$
\cite{fukuda}. The analysis of Superkamiokande yields roughly the
same mixing angle but a lower value of $\Delta m^2 \simeq
0.001 \ eV^2$ \cite{superk,gg}. 
Since both $\nu_\mu \leftrightarrow \nu_e$ and      
$\nu_\mu \leftrightarrow \nu_\tau$ oscillations give equally good 
allowed regions, one must look for ways of distinguishing which 
type of oscillations are the cause of the atmospheric neutrino
anomaly. 

Recently Vissani and Smirnov (VS) \cite{vismir} proposed that the 
single pion events in the energy range 0.5 GeV - 1.5 GeV can be used
to distinguish between different types of oscillations. Such events
induced by neutral current (NC) contain a $\pi^0$ whereas the events
induced by charged current (CC) contain a $\pi^+$ or $\pi^-$. The 
$\pi^\pm$ are detected as sharp rings, whereas the 
$\pi^0$
decays into two photons which are detected as two diffuse rings.
Kamiokande has already observed such $\pi^0$'s in their detector
by considering all events with two diffuse rings and by selecting
those events with invariant mass in the range 90 MeV - 180 MeV
\cite{fukuda1}.
The CC events due to $\nu_e$ contain an $e^\mp$ in addition to a 
$\pi^\pm$ and those due to $\nu_\mu$ contain a $\mu^\mp$. Thus one
can distinguish an NC event (two difuse rings with the invariant
mass of the rings around $m_\pi$) from a $\nu_e$ CC event (one 
diffuse ring due to $e^\mp$ and one sharp ring due to $\pi^\pm$)
and from a $\nu_\mu$ CC event (two sharp rings from $\mu^\mp$ and $\pi^\pm$).
The two ratios, NC events to $\nu_e$ CC events and NC events to $\nu_\mu$
CC events, are free from the uncertainties of the theoretical flux
calculations. The ratio of the cross sections of the NC to CC events 
in single pion production were measured previously \cite{takita} and
the uncertainties in them are about $15 \%$. Hence a measurement of the 
above ratios can help in distinguishing the different types of 
oscillations which are currently relevant for atmospheric neutrino
anomaly.

In this paper we analyze the atmospheric neutrino problem in the 
framework of oscillations between three active flavors. We will assume 
that one of the masses is much greater than the other two. We fix the 
smaller of the mass-squared differences using the solar neutrino data
\cite{nmru,flm}. Then the larger mass-squared difference and two of
the mixing angles are relevant for the atmospheric neutrino problem
\cite{jim}. The double ratio $R$ defined in equation(1), 
along with the zenith angle dependent
multi-GeV data, was analyzed in this framework previously \cite{flms,nru}.
Here we analyze the ratio 
of CC to NC events for one-pion production
in the framework of three active neutrino oscillations
and see how they can distinguish between different 
regions of parameter space. The NC event rate is  unaffected by the 
oscillations while the CC event rate is affected by the 
oscillations and hence depend on the neutrino 
parameters. In section II we give the basic theory for the 
calculation of the one-pion event rates in 
the three-flavor oscillation scenario. This is followed by the 
discussion of the experimentally
measurable quantities in section III and a summary is given in section IV. 

\section{THEORY }

In a three flavor scheme, the weak eigenstates 
$\nu_{\alpha}$ are related to the mass eigenstates 
$\nu_{i}$ through a 3$\times$3 unitary matrix $U$ as 
\begin{equation}
\nu_{\alpha} = \sum_{i}U_{\alpha i}\nu_{i}.
\end{equation}
$U$ can be written as
\begin{equation}
U =  U^{23}(\psi)\times U^{phase}\times U^{13}(\phi)
\times U^{12}(\omega), 
\end{equation}
where, $U^{ij}(\theta_{ij})$ is the two flavor mixing 
matrix between the $i$th and $j$th mass 
eigenstates with mixing angles $\theta_{ij}$. 
We assume CP invariance and set $U^{phase} = I$.
For the neutrinos that propagate through matter, the CC 
interaction between $\nu_{e}$ and $e$ induces an 
effective mass term for the $\nu_{e}$ which is of the form $A=
2\sqrt{2}G_{F}N_{e}E$, where $N_{e}$ is the electron number 
density and $E$ is neutrino energy. Single-pion 
production maximally occurs for neutrinos within the energy range of 
0.5 GeV - 1.5 GeV. 
For this energy region, matter effects in the earth can be 
neglected  since $A<{10}^{-4}~eV^2$ and  
$\delta_{31} \simeq {10}^{-3}~eV^2$, as given by Superkamiokande
analysis \cite{superk,gg}. This choice for $\delta_{31}$ is
consistent with the mass heirarchy assumption under which
both the atmospheric and solar neutrino anamalies can be explained. 
Hence, here we have the simple case of vacuum oscillations.  
The vacuum oscillation probability $P_{\alpha\beta}$ for a 
neutrino with a flavor $\alpha$
to oscillate in to a flavor $\beta$ can be given by  
\begin{eqnarray}
P_{\alpha\beta} = U^{2}_{\alpha 1}U^{2}_{\beta 1} + 
U^{2}_{\alpha 2}U^{2}_{\beta 2} + 
U^{2}_{\alpha 3}U^{2}_{\beta 3} + 2U_{\alpha 1}U_{\alpha 2}U_{\beta 1}
U_{\beta 2}cos \left(\frac{2.53x\delta_{21}}{E} \right) +\nonumber \\ 
2U_{\alpha 1}U_{\alpha 3}U_{\beta 1}U_{\beta 3}
cos \left( \frac{2.53x\delta_{31}}{E} \right) 
+ 2U_{\alpha 3}U_{\alpha 2}U_{\beta 3}U_{\beta 2}
cos \left( \frac{2.53x\delta_{32}}{E} \right).
\end{eqnarray}
Here, $E$ is the energy of the neutrino in GeV, 
$x$ is the distance travelled by the neutrino in 
kilometers and $\delta_{ij}$ is $m^2_{i}-m^2_{j}$ in $eV^2$. 
We assume the mass 
heirarchy ($\delta_{21}$ $\ll ~ \delta_{32}$ $\simeq$ 
$\delta_{31}$) as suggested by the variuos analyses
of the solar and atmospheric neutrino problems
\cite{fukuda,flms,nru,nmru,flm,jim,halg} and take 
$\delta_{21} ~ \simeq ~0$, which allows us 
to put the oscillating term involving $\delta_{21}$ as unity. So we get
\begin{equation}
P_{\alpha\beta} = (U_{\alpha 1}U_{\beta 1} + 
U_{\alpha 2}U_{\beta 2})^2 + (U_{\alpha 3}U_{\beta 3})^2
+ 2(U_{\alpha 1}U_{\beta 1} + U_{\alpha 2}U_{\beta 2})U_{\alpha 3}
U_{\beta 3}cos \left( \frac{2.53\delta_{31}x}{E} \right).
\end{equation}
Since we are considering vacuum oscillations with CP invariance, 
the oscillation probability for 
antineutrinos is the same as that for neutrinos.
 
  The various one-pion events arising from $NC$ and $CC$ 
interaction are listed below:
\begin{eqnarray}
\nu_{l} N & \rightarrow &
\nu_{l} \pi^0 N,~~~~~~~~~l = e, \mu, \tau,  
\label{eq:NC} \\
\nu_{l} N & \rightarrow &  
l \pi^0 N^{\prime},~~~~~~~~l = e, \mu, 
\label{eq:nuCC} \\ 
\nu_{e} N & \rightarrow & e^{-} \pi^{+} N, 
\label{eq:eCC} \\ 
\nu_{\mu} N & \rightarrow & \mu^{-}\pi^{+} N,  
\label{eq:muCC}
\end{eqnarray}
where, $N$ and $N^{\prime}$ refer to the nucleons. As we are dealing with sub-GeV neutrinos, 
$\nu_\tau$ do not excite $CC$ processes.
For every neutrino reaction, we consider the corresponding      
anti-neutrino reaction also. 

   It is difficult to calculate 
the zenith-angle dependent fluxes for sub-GeV neutrinos 
\cite{honda,agarwal} and further the scattering angle of the final state 
charged lepton for neutrinos in the sub-GeV range 
can be as large as ${60}^o$ and thus directionality is lost in the detection process. 
Moreover, the event rates for the one-pion events
that we consider are not very large. Hence we will not consider 
zenith angle dependence and will average over the
zenith angle, or equivalently, over $x$, the distance travelled 
by the neutrinos in its allowed range (20 Km -13000 Km).

   The NC rate $N^{NC}$ (with $\pi^0$ as the detectable final state
particle) is given by 
\begin{equation}
N^{NC} =  
\int^{1.5 GeV}_{0.5 GeV} dE  \left[\frac{d \Phi_e(E)}{dE} +
\frac{d \Phi_\mu(E)}{dE} \right] \sigma^{NC}(E)~\epsilon_{\pi^0},
\label{eq:NNC}
\end{equation} 
where, $\sigma^{NC}$ is the scattering cross section for 
the reaction of Eq.~({\ref{eq:NC})
and $\epsilon_{\pi^0}$ is the $\pi^0$ detection efficiency 
which is 0.77 for the 
energy range of interest. $N^{NC}$ being flavor blind remains 
unaffected by oscillations. The CC rates 
arising from processes (7-9) can be written as 
\begin{eqnarray} 
N^{CC}_{\mu^-\pi^+} & = & 
\frac{1}{12980} \int^{1.5 GeV}_{0.5 GeV} dE  
\int^{13000 Km}_{20 Km} dx \left[\frac{d \Phi_\mu(E)}{dE} P_{\mu\mu} (E,x) +
\frac{d \Phi_e(E)}{dE} P_{e\mu} (E,x) \right] \sigma_{\mu^-}^{\pi^+}(E) 
 \epsilon_{\pi^+} \epsilon_{\mu^-},
\label{eq:NmuCC} \\ 
N^{CC}_{e^-\pi^+} & = &  
\frac{1}{12980} \int^{1.5 GeV}_{0.5 GeV} dE  
\int^{13000 Km}_{20 Km} dx \left[\frac{d \Phi_e(E)}{dE} P_{ee} (E,x) + 
\frac{d \Phi_\mu}{dE} (E) P_{\mu e} (E,x) \right] \sigma_{e^-}^{\pi^+}(E)
\epsilon_{\pi^+}~\epsilon_{e^-}.  
\label{eq:NeCC} \\ 
N^{CC}_{\mu^-\pi^0} & = & 
\frac{1}{12980} \int^{1.5 GeV}_{0.5 GeV} dE 
\int^{13000 Km}_{20 Km} dx \left[\frac{d \Phi_\mu(E)}{dE} P_{\mu\mu} (E,x) +
\frac{d \Phi_e(E)}{dE} P_{e\mu} (E,x) \right] \sigma_{\mu^-}^{\pi^0}(E) 
\epsilon_{\pi^0} \epsilon_{\mu^-},
\label{eq:Nmupi0} \\ 
N^{CC}_{e^-\pi^0} & = &  
\frac{1}{12980} \int^{1.5 GeV}_{0.5 GeV} dE  
\int^{13000 Km}_{20 Km} dx \left[\frac{d \Phi_e(E)}{dE} P_{ee} (E,x) + 
\frac{d \Phi_\mu}{dE} (E) P_{\mu e} (E,x) \right] \sigma_{e^-}^{\pi^0}(E)
\epsilon_{\pi^0}~\epsilon_{e^-}.  
\label{eq:Nepi0} 
\end{eqnarray}
In each of the above equations
the events due to antineutrinos are added to those due to
neutrinos. Here, $\epsilon_{e^+}$, $\epsilon_{\mu^+}$ and  
$\epsilon_{\pi^-}$ are the electron, muon 
and the charged pion detection efficiencies respectively, which in 
principle are energy dependent quantities. 
However, for the limited energy
range of sub-GeV neutrinos, they can be taken to be energy 
independent and to be close to unity \cite{hirata,fukuda1}.
The differential fluxes for electron and muon neutrinos, 
$d \Phi_e /dE$ and $d \Phi_\mu/dE$, are taken from \cite{honda},
$\nu_\mu$-nucleon CC cross section is taken from \cite{takita} and       
$\nu_e$-nucleon CC cross section is obtained by the approximation
$\sigma_{e^-}^{\pi^+} (E) \simeq \sigma_{\mu^-}^{\pi^+} 
(E + 100~MeV)$ \cite{hirata}.
The NC cross sections for neutrino energy range 0.5 GeV - 1.5 GeV
were taken from \cite{faissner,lee,takita}. 
The corresponding rates in the absence of oscillations can be 
obtained by setting the survival probability
$P_{\alpha\alpha}$ to unity and transition probabilities 
$P_{\alpha\beta}$ ($\alpha \neq \beta$) to zero.

\section{ EXPERIMENTALLY MEASURABLE QUANTITIES} 

 From the processes listed in Eqs.~({\ref{eq:NC}) - ~({\ref{eq:muCC}) 
we see that a $\pi^0$ is produced 
both in the neutral current process defined
in Eq.~({\ref{eq:NC}) and in the charged current process defined 
in Eq.~({\ref{eq:nuCC}).  The latter
process, in general produces three rings (two from the $\pi^0$ decay 
and one from the charged lepton)
whereas the former produces only two rings. As discussed in the 
introduction, the charged current processes
in Eq.~({\ref{eq:eCC}) and~({\ref{eq:muCC}) also produce two-ring 
events. Hence, by choosing only
two-ring events we can discriminate against the charged current 
events of the type given in
Eq.~({\ref{eq:nuCC}). However, there is a siginificant probability 
of about 0.17 when the two photons
from the $\pi^0$ decay cannot be resolved. If the $\pi^0$ produced 
in $\nu_e N \rightarrow e \pi^0 N'$
is not resolved into two photons, then we see two diffuse rings 
in this process also. Such events can be usually 
rejected 
because in general their invariant mass will not be near $m_\pi$. 
However, if the $\pi^0$ produced in $\nu_\mu N
\rightarrow \mu \pi^0 N'$ is unresolved, we have a signal of one 
diffuse and one sharp ring and this process
forms a background to the reaction in Eq.~({\ref{eq:eCC}). 
Its effect in all the experimental
observables must be properly taken into account.  
An additional process we take into account is the 
production of two charged pions via the neutral current process 
\begin{equation} 
\nu_l N \rightarrow \nu_l~\pi^+\pi^- N. \label{eq:2pi} 
\end{equation} 
This process produces two sharp rings and mimics the reaction given
in equation~(\ref{eq:muCC}). We take its rate to be about $10 \%$ 
of the rate of reaction in
equation~(\ref{eq:muCC}) \cite{vismir}. 

Let us define the following experimentally measurable quantities:
\begin{enumerate}
\item
$N_{DD}$: The number of events with two diffuse rings.
\item
$N_{\pi^0}$: The number of events with two diffuse rings whose 
invariant mass is in the range 90 MeV - 180 MeV.
\item
$N_{DS}$: The number of events with one diffuse and one sharp ring.
\item
$N_{SS}$: The number of events with two sharp rings.
\end{enumerate}
If the detector were ideal and all the particles can be exactly 
identified, the relations between experimentally 
measured quantities and theoretically calculable quantities would
be $N_{\pi^0} = N^{NC}$, $N_{DS} = N^{CC}_{e^- \pi^+}$ and 
$N_{SS} = N^{CC}_{\mu^- \pi^+}$. The charged particles can be 
identified with good efficiency but the efficiency of identifying
$\pi^0$ events from $N_{DD}$ is only $0.77$. There is a $0.17$ 
probability that the two photons from $\pi^0$ decay cannot be 
resolved and the $\pi^0$ appears as a single diffuse ring. Thus the
CC events with a $\pi^0$ in the final state lead to contributions to
$N_{DS}$ and $N_{SS}$. Hence, the modified relations between the experimental
and theoretical quantities are
\begin{eqnarray}
N_{\pi^0} & = & 0.77 N^{NC} + 0.17 \times 0.4 N^{CC}_{e^- \pi^0} 
\nonumber \\
N_{DS} & = & N^{CC}_{e^- \pi^+} + 0.17 N^{CC}_{\mu^- \pi^0} \nonumber \\
N_{SS} & = & N^{CC}_{\mu^- \pi^+} + N^{NC}_{\pi^+ \pi^-}. 
\end{eqnarray}
In writing the above equations, we added the contribution of
$\nu_\mu N \rightarrow \mu^- \pi^0 N'$ to $N_{DS}$ due to the 
$17 \%$ $\pi^0$ misidentification probability and the small
contribution to $N_{SS}$ due to two charged pion production.
We also took into account the contribution of 
$\nu_e N \rightarrow e^- \pi^0 N'$ to $N_{\pi^0}$ due to the  
$17 \%$ $\pi^0$ misidentification probability. Assuming that  
the directions of the electron and the misidentified $\pi^0$ 
are randomly distributed, 
we estimate that only about $40 \%$ of these events survive 
the cut on the invariant mass of the two diffuse rings. Hence
the factor $0.4$ in the correction to the expression for $N_{\pi^0}$.

 From the experimentally measurable quantities, we define the 
following three ratios
\begin{eqnarray} 
R_1 & = &  \frac{N_{DS}}{N_{\pi^0}},
\label{eq:defR1} \\
R_2 & = &  \frac{N_{SS}}{N_{\pi^0}},
\label{eq:defR2} \\
R_3 & = &  \frac{N_{DS}+N_{SS}}{N_{\pi^0}}.
\label{eq:defR3} 
\end{eqnarray}
Presently the data on one pion events in the atmospheric neutrino
experiments is very limited and the errors are quite large. However,
we will assume that the large statistics of Superkamiokande will enable 
it to measure all experimental quantities to an accuracy of better
than $10 \%$. Turning now to theoretical uncertainties, the single
largest source of error occurs in the ratio \cite{faissner} 
\begin{equation}
\frac{ \sigma (\nu_\mu p \rightarrow \nu_\mu p \pi^0) +  
\sigma (\nu_\mu n \rightarrow \nu_\mu n \pi^0)}{
2 \sigma (\nu_\mu n \rightarrow \mu^- p \pi^0)} = 0.47 \pm 0.06.
\end{equation}
This, when combined with other errors, leads to a minimum theoretical
uncertainty of about $15 \%$ in the predictions of $R_1$ and $R_2$.
Unfortunately, this uncertainty cannot be reduced and will be part
of all future predictions. Still, even with this handicap, we find that
the ratios $R_1$ and $R_2$ can be used to distinguish between different
types of neutrino oscillations.

For the atmospheric neutrino oscillations, based on the mass hierarchy 
assumption and using CP
invariance, the three relevant parameters are $\phi$, $\psi$ and 
$\delta_{31}$.  We choose the following
ranges for the mixing angles: 
\begin{itemize} 
\item $0\leq\phi\leq50^o$. This comes from the analysis of
the solar neutrino anomaly. For this region of $\phi$, 
there exist allowed values of $\omega$ and
$\delta_{21}$ which can account for the various solar neutrino data 
\cite{nmru,flm}.  
\item $30\leq\psi\leq90^o$. This comes from our analysis of the 
zenith angle dependent single ring events of
Kamiokande data \cite{nru} on atmospheric neutrinos.  
\end{itemize} 
For definiteness, we fix $\delta_{31}
= 0.001~eV^2$, which is the central value favored by Superkamiokande 
analysis of the single ring
atmospheric data \cite{superk,gg}. However, the results do not depend on
$\delta_{31}$ significantly. The reason for this is that we are 
averaging over the distance in
calculating the number of events and the dependence on 
$\delta_{31}$ is lost during this averaging.
We will try to explore how useful the ratios $R_1$
and $R_2$ are in distinguishing different regions of the mixing
angles. We note that without oscillations (i.e., $\phi ~=~0$ and 
$\psi ~=~0$), these ratios have the values
$R_1^{no} = 2.6$ and $R_2^{no} = 3.4$. We plot the values of $R_1$ and 
$R_2$ as 
functions of the mixing angle $\psi$ for different values of $\phi$ in 
figures 1 and 2 resectively. From 
these we note that for large values of mixing
angles, $R_1$ and $R_2$ differ significantly from their no oscillation values.

Let us consider three
particular cases of maximal oscillations:
\begin{enumerate}
\item
$\nu_\mu \leftrightarrow \nu_e$ oscillations: In this case,
$\psi = 90^o$ and $\phi = 45^o$. Since flux of $\nu_\mu$ 
is larger than that of $\nu_e$, the total flux of $\nu_e$
increases and that of $\nu_\mu$ decreases. Hence $R_1$ increases
and $R_2$ decreases. For this case, $R_1 = 3.8$ and $R_2 = 2.5$.
Clearly the change in $R_1$ and $R_2$ is more than the $15 \%$
uncertainty in the theoretical prediction and one can distinguish
this case from that of no oscillations.
\item
$\nu_\mu \leftrightarrow \nu_\tau$ oscillations: In this case,
$\psi = 45^o$ and $\phi = 0^o$. Here, the flux of $\nu_\mu$ 
goes down and that of $\nu_e$ is unaffected. Hence $R_1$ does 
not change and $R_2$ decreases. Numerically we have, $R_1 = 2.6$
and $R_2 = 1.9$. These numbers are significantly different from
those of the previous case as well as that of no oscillations.
Hence the $15 \%$ theoretical uncertainty is no bar in distinguishing
this case.
\item
Maximal three flavor oscillations: Here all the three flavours mix
maximally with one another. This occurs for $\psi = 45^o$ and 
$\phi \simeq 35^o$. For this case, $R_1 = 2.6$ and $R_2 = 2.2$.
Thus the values of $R_1$ and $R_2$ together allow us distinguish it 
from the no oscillation case as well as the two previous cases.
\end{enumerate}

Finally, let us consider the ratio $R_3$ which is the sum of $R_1$ 
and $R_2$ because already one can extract a value for it 
from Kamiokande data. The experimental uncertainty is quite large but we will  
derive the constraints on the neutrino parameter space from the
existing data. From \cite{fukuda1}, we estimate $R_3 = 4.2 \pm 0.7$.
The experimental error is about $16 \%$. To this, we add the 
$15 \%$ theoretical error in quadrature. Thus we have 
$R_3 = 4.2 \pm 1.0$. Fig. 3 shows the plots of $R_3$ as functions
of the angle $\psi$ for different values of $\phi$.
We note that the upper limit $R_3 \leq 5.2$ imposes the constraint
$\psi \leq 70^o$. Hence pure $\nu_\mu \leftrightarrow \nu_e$ 
oscillations are disfavoured at $1 \sigma$ level. 
Figure~4 shows the allowed region in $\phi-\psi$ plane
by the Kamiokande constraint on $R_3$. Already, with the
very limited data available, the ratios of 1-pion events yield 
useful constraints on the allowed neutrino parameter space.

\section{SUMMARY AND DISCUSSION}

The atmospheric neutrino experiments have already given us 
valuable information on the neutrino
parameters. So far the bulk of the analysis of the experimental 
data has been restricted to 
the single ring events. With the accumulation of more abundant 
data in the existing and the upcoming 
detectors, analysis of two-ring events is also likely to be 
taken up in the future. We have used the 
framework of the three-flavor oscillations, to suggest such an 
analysis. We have shown that independent 
information on the parameters $\phi$ and $\psi$ can be obtained.

  Recently the results of CHOOZ experiment led to strong 
  constraint on the mixing angle $\phi \leq 12.5^o$
\cite{chooz,nruchooz,bilenky}.  This in turn implies that the 
atmospheric neutrino anomaly is almost due
to $\nu_\mu \leftrightarrow \nu_\tau$ oscillations. Current 
Superkamiokande data on single ring events
also favors this explanation. Data on two ring events can provide 
a dramatic confirmation of this
scenario. 
\vskip 2cm

\begin{center} 
{\bf Acknowledgements} 
\end{center}

We thank M. V. N. Murthy, Rahul Sinha, Shashi Kant Dugad and 
Mohan Narayan for useful 
conversations. One of us (Balaji) 
thanks the EHEP, TIFR in Mumbai, and S. S. Rao of IMSc Chennai.

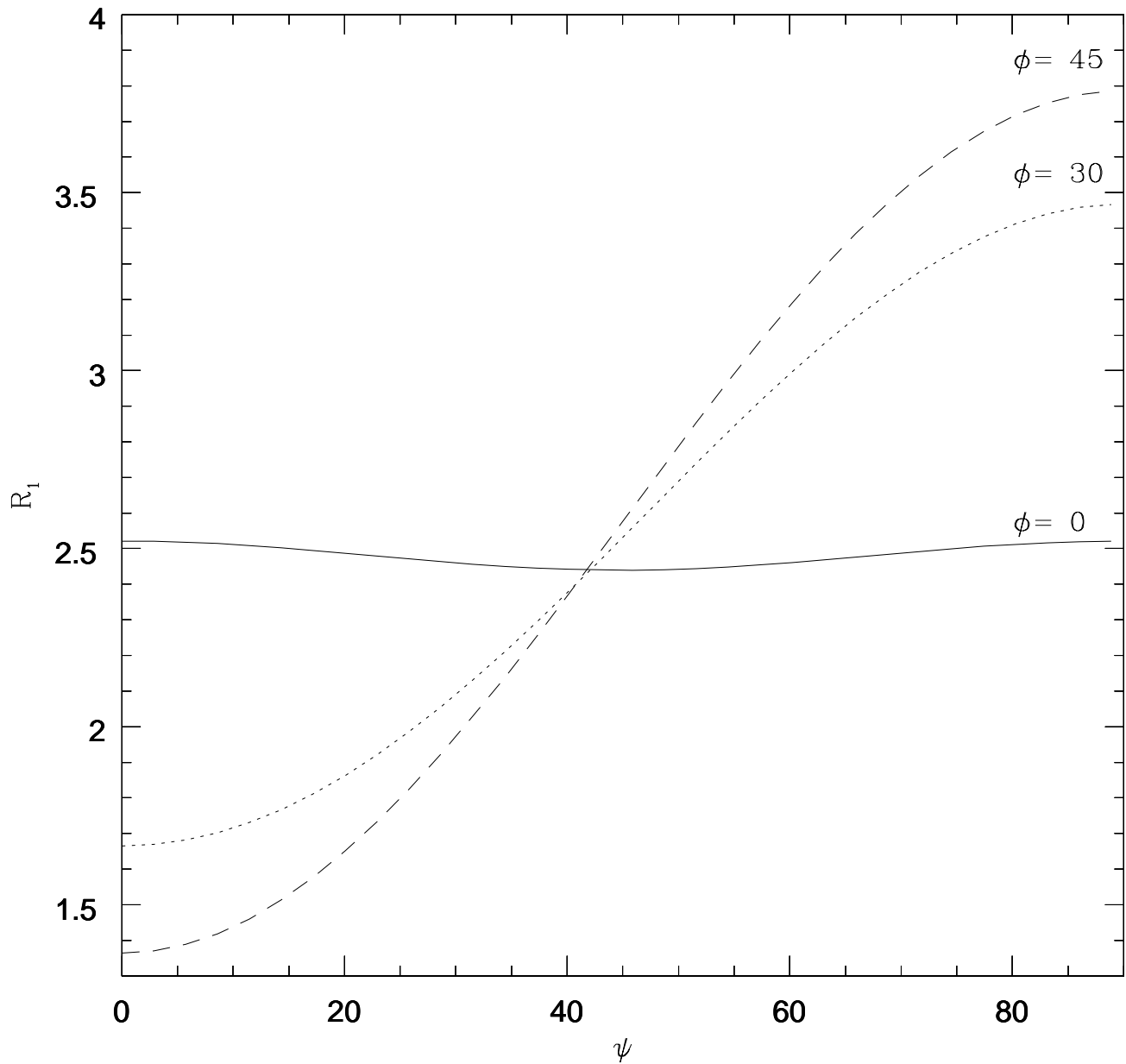
\begin{figure}
\caption{Plot of $R_1$ vs $\psi$ for $\delta_{31} = 0.001 \ eV^2$
and $\phi = 0$ (continuous line), $\phi = 30^0$ (dotted line) and 
$\phi = 45^o$ (dashed line).
}\label{Fig. 1}
\end{figure}

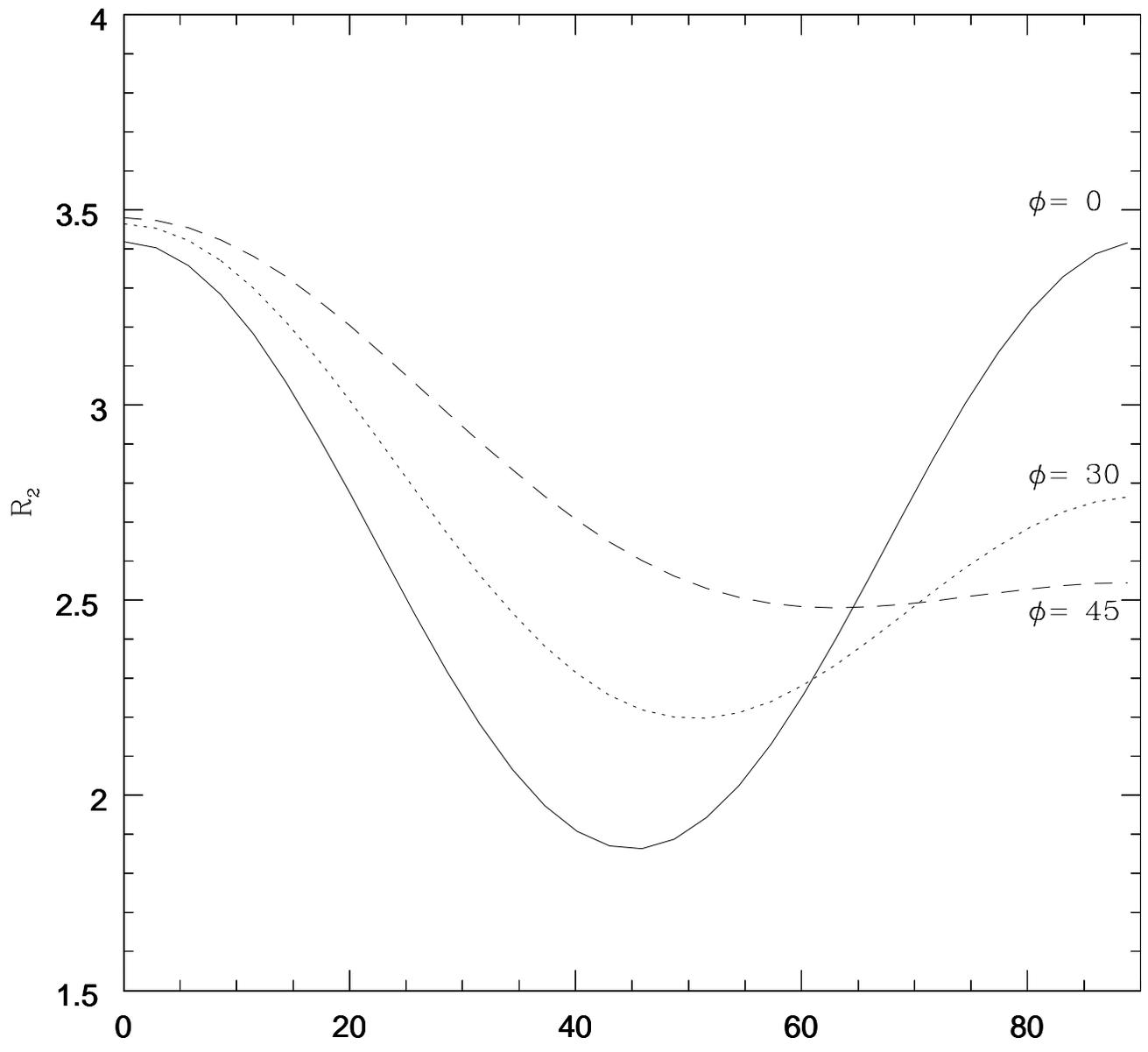
\begin{figure}
\caption{Plot of $R_2$ vs $\psi$ for $\delta_{31} = 0.001 \ eV^2$
and $\phi = 0$ (continuous line), $\phi = 30^0$ (dotted line) and 
$\phi = 45^o$ (dashed line).
}\label{Fig. 2}
\end{figure}

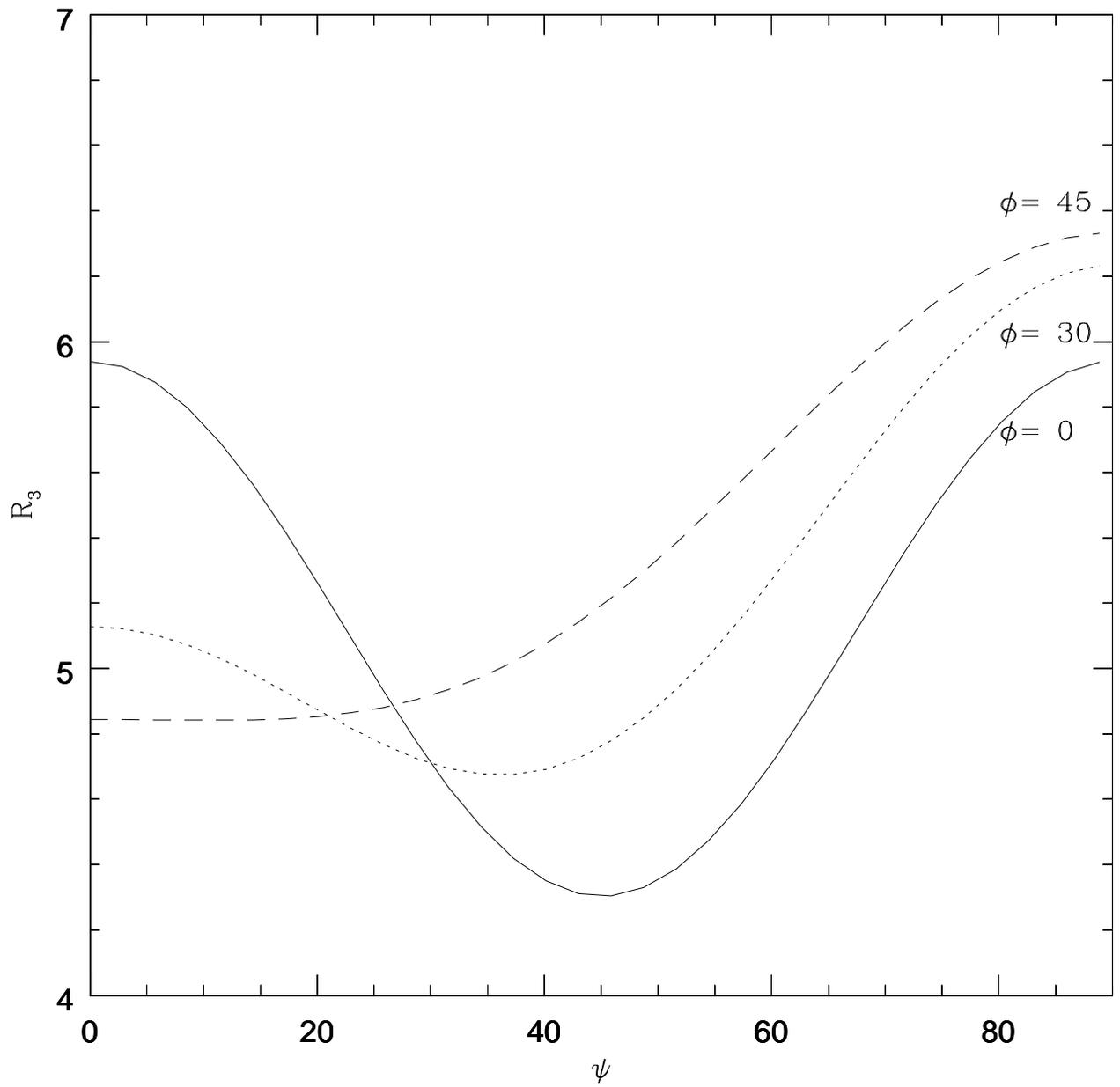
\begin{figure}
\caption{Plot of $R_3$ vs $\psi$ for $\delta_{31} = 0.001 \ eV^2$
and $\phi = 0$ (continuous line), $\phi = 30^0$ (dotted line) and 
$\phi = 45^o$ (dashed line).
}\label{Fig. 3}
\end{figure}

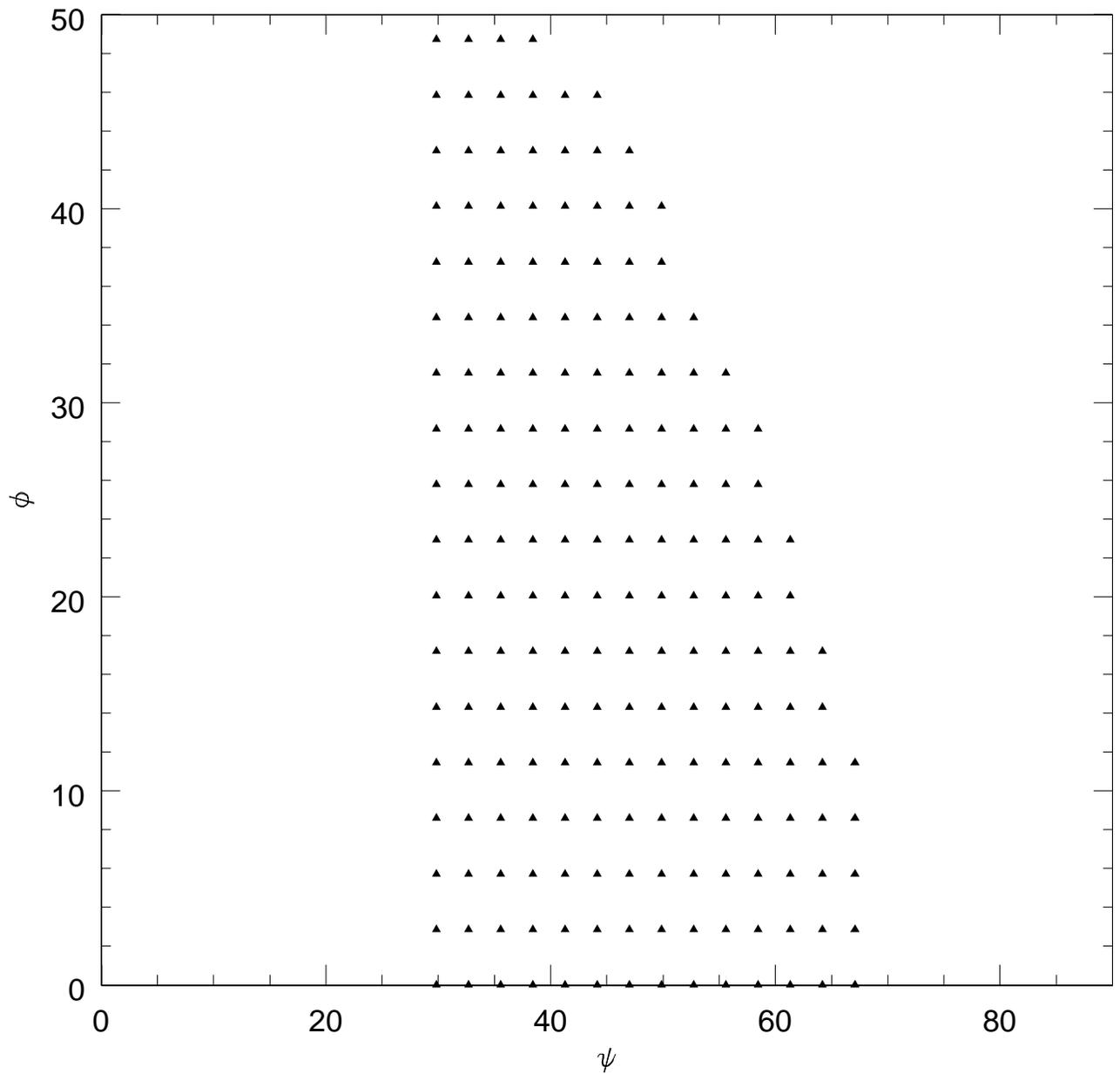
\begin{figure}
\caption{Region in $\phi-\psi$ plane allowed by the Kamiokande 
constraint on $R_3$.~ ~ ~ ~ ~ ~ ~ ~ ~ ~ ~ ~ ~ 
}\label{Fig. 4}
\end{figure}

\end{document}